\documentclass{Interspeech2024}
\UseRawInputEncoding

\interspeechcameraready

\title{Spontaneous Style Text-to-Speech Synthesis with Controllable Spontaneous Behaviors Based on Language Models}

\name[]{Weiqin Li$^{1,*}$, Peiji Yang$^{2}$, Yicheng Zhong$^{2}$, Yixuan Zhou$^{1}$, \\Zhisheng Wang$^{2}$, Zhiyong Wu$^{1,3,\dagger}$, Xixin Wu$^{3}$, Helen Meng$^{3}$}{}

\address{
  $^1$ Shenzhen International Graduate School, Tsinghua University, Shenzhen, China\\
  $^2$ Tencent, Shenzhen, China\\
  $^3$ The Chinese University of Hong Kong, Hong Kong SAR, China}
\email{\{lwq22@mails, zywu@sz\}.tsinghua.edu.cn, \{peijiyang, ajaxzhong, plorywang\}@tencent.com\thanks{* This work is done during Weiqin Li's internship at Tencent.}\thanks{$\dagger$ Corresponding author.}}

\newcommand{\red}[1]{\textcolor{red}{#1}}
\begin{document}

\maketitle
\begin{abstract}
Spontaneous style speech synthesis, which aims to generate human-like speech, often encounters challenges due to the scarcity of high-quality data and limitations in model capabilities.
Recent language model-based TTS systems can be trained on large, diverse, and low-quality speech datasets, resulting in highly natural synthesized speech.
However, they are limited by the difficulty of simulating various spontaneous behaviors and capturing prosody variations in spontaneous speech.
In this paper, we propose a novel spontaneous speech synthesis system based on language models.
We systematically categorize and uniformly model diverse spontaneous behaviors.
Moreover, fine-grained prosody modeling is introduced to enhance the model's ability to capture subtle prosody variations in spontaneous speech.
Experimental results show that our proposed method significantly outperforms the baseline methods in terms of prosody naturalness and spontaneous behavior naturalness.

\end{abstract}
\section{Introduction}
Text-to-speech (TTS) aims to synthesize intelligible and natural speech from text~\cite{survey}.
With the development of deep learning, existing TTS models are able to generate highly expressive speech~\cite{fastspeech2,tacotron2,vits}. 
Nonetheless, the synthesis of speech with a spontaneous style, which usually occurs in daily conversations, talk shows, and podcasts, has not been well studied.


Several researchers have explored spontaneous speech synthesis 
by conducting high-quality and well-annotated spontaneous speech corpus to train neural TTS models~\cite{converTTS}. 
Incorporating
explicit labels to model and control spontaneous behaviors in speech have been proven to 
be effective in producing more human-like speech~\cite{controllabelSpon}.
However, they typically focused on 
only a few
specific types of spontaneous behaviors, such as
filled pause~\cite{Adaspeech3,lwqis23}, breathing~\cite{breathingSpontaneous}, interjections~\cite{interjections}, laughter~\cite{latenStyleSpontaneous}, and creaky phonation~\cite{creaky}, 
neglecting the wide diversity of spontaneous behaviors,
which limits the expressiveness of generated speech.
Some studies have extracted global and local speaking style representations as latent spontaneous prosody features, and predicted them from text in an implicit manner~\cite{latenStyleSpontaneous,spontts}.
But capturing the diverse and complex prosody characteristics of spontaneous speech poses a significant challenge.
Additionally, relying solely on text information, without considering the cues from spontaneous behaviors, also hinders the accurate prediction of prosody features.

Previous works on spontaneous speech synthesis are trained on limited data, leading to a discernible disparity in the naturalness of the synthesized speech compared to real human speech.
A data augmentation method based on semi-supervised pre-training has been proposed to utilize low-quality spontaneous style corpus, 
neglecting the impact of noise from low-quality data on the synthesized speech~\cite{lwqis23}.
Inspired by the advancements in text language models, recent TTS systems~\cite{valle,vallex,spearTTS,audioLM} encode speech waveforms into discrete tokens with neural audio codecs~\cite{hneural,soundstream} as an intermediate representation, and model them with the paradigm of prompt-based language models.
These language model (LM)-based TTS models can be trained on large, diverse, and low-quality speech datasets. 
Besides, benefiting from the in-context learning ability of LM,
these models are able to capture local and long-range dependencies in speech while also acquiring a strong semantic understanding, resulting in a substantial enhancement in the diversity and naturalness of the generated speech.

However, 
current LM-based TTS models
primarily focus on voice cloning and have not been well explored for generating human-like, authentic spontaneous speech.
Although these models can learn some spontaneous expressions from a large amount of speech data, there are still two main challenges in achieving high-quality spontaneous speech synthesis:
1)
The diverse spontaneous behaviors that distinguish spontaneous speech from read speech~\cite{breathingSpontaneous,SynthesisingUncertainty}
necessitate effective modeling and control;
2) The intrinsically diverse prosody variations in spontaneous speech are difficult for TTS models to adequately capture,
which limits the performance of prosody details in generated speech.
In this paper, 
we propose a novel spontaneous speech synthesis system based on language models that can learn powerful semantic understanding and diverse speech expressions effectively from a massive amount of data.
To categorize and control various spontaneous speech patterns in human speech,
we have identified 19 spontaneous behaviors based on linguistic characteristics and explicitly modeled them by providing behavior labels and syntactic information, 
which is the most comprehensive work
to the best of our knowledge.
In addition, we introduce fine-grained spontaneous prosody representations and utilize context, spontaneous behavior labels and linguistic information to predict them.
Experimental results show that our proposed method enables the LM-based TTS model to synthesize more natural spontaneous style speech.
Explicit label control enhances the model's ability to simulate spontaneous behaviors, while the addition of spontaneous prosody modeling significantly improves the naturalness of synthesized speech.
\section{Methodology}
\begin{figure}[!tb]
	\centering
	\includegraphics[width=1 \linewidth, height=1.4\linewidth]{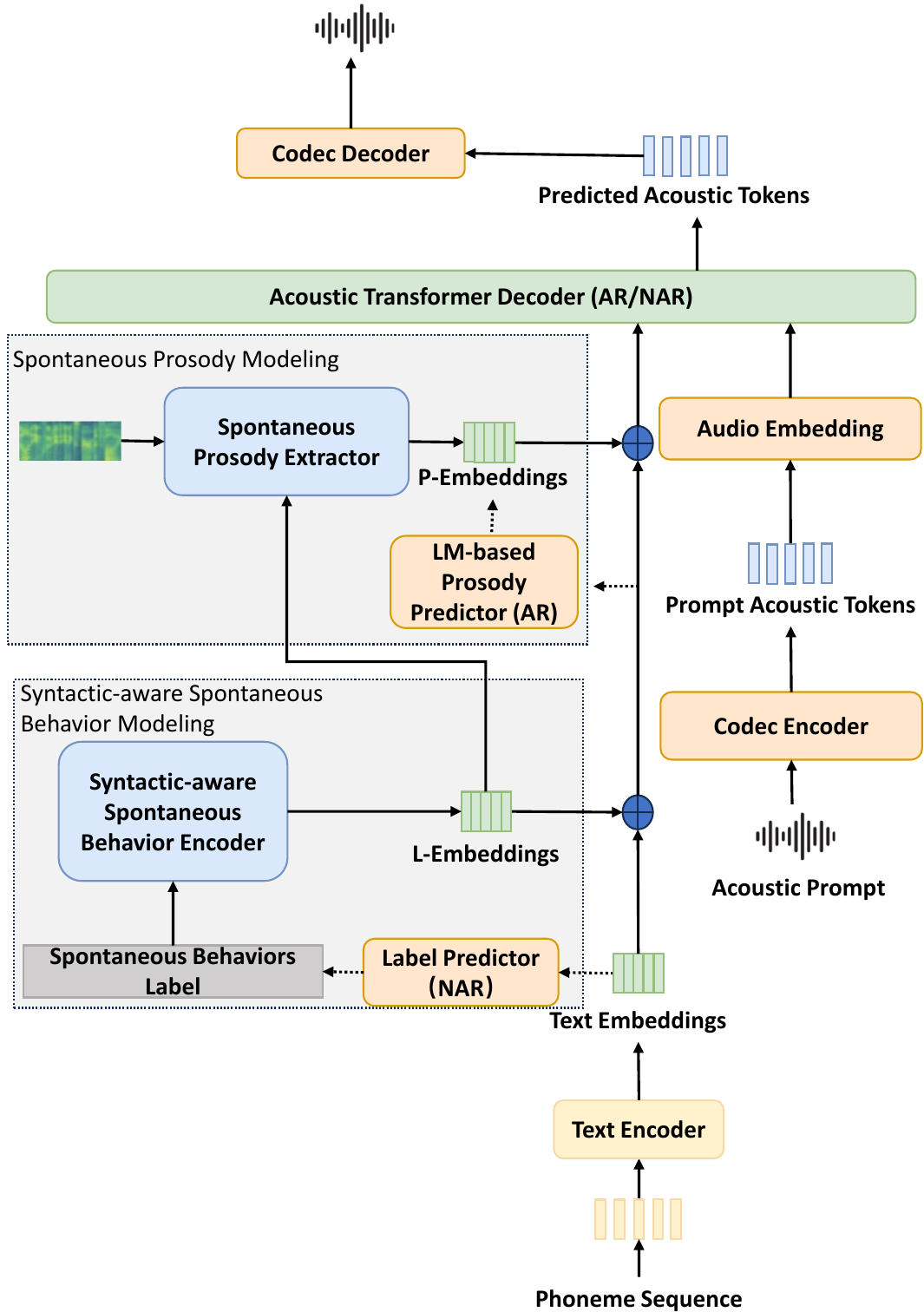}
	\caption{The architecture of our proposed model. Label predictor and prosody predictor outputs are used for inference.}
	\label{fig:struct}
        \vspace{-0.6cm}
\end{figure}

The architecture of our proposed model is illustrated in Figure. \ref{fig:struct}.
We use an acoustic decoder based on VALL-E~\cite{valle} as the backbone, with text and 
acoustic 
prompts as conditions. 
To model the spontaneous behaviors, we introduce a syntactic-aware spontaneous behavior encoder, and use a label predictor to predict spontaneous labels from text embeddings.
Besides, a spontaneous prosody extractor and a LM-based prosody predictor are used to guide the fine-grained spontaneous prosody modeling, for enhancing the expressiveness of the synthesised speech.
In the following subsections, we describe the major parts of the model, and the methods of pre-training and fine-tuning.

\vspace{-0.2cm}
\subsection{Backbone Framework}
\vspace{-0.1cm}

Inspired by the success of LM-based TTS models, we adopt VALL-E~\cite{valle} as the model backbone, as shown in Figure.~\ref{fig:struct}.
It consists of a text encoder, 
an audio embedding,
an autoregressive (AR) transformer decoder, a non-autoregressive (NAR) transformer decoder,
and a well-trained neural audio codec.
The data flow and training method of the backbone model are the same as VALL-E.


We conduct pre-training of the backbone on a large-scale dataset and subsequently fine-tune it on a spontaneous style corpus.
This process results in a decoder's feature space that more closely aligns with human performance.
Since the training process of the AR transformer decoder does not use explicit prompts, to ensure consistency between training and inference, we concatenate each audio clip with another audio segment during the fine-tuning stage for data augmentation.

\vspace{-0.6cm}
\subsection{Analysis and Taxonomy of Spontaneous Behaviors}
\vspace{-0.1cm}
Spontaneous behaviors refer to the unusual durations or pitch excursions produced by paralinguistic phenomena in speech that are not fully predictable from syntax.
Based on previous researches, we classify the diverse spontaneous behaviors in Mandarin into three major categories: disfluency, interjections, and non-speech sounds~\cite{spontaneousClassify}:

\textbf{(1) Disfluency:}
Disfluency occurs when speech becomes incoherent due to hesitation, nervousness, thinking, or emphasis.
It includes several types, such as filled pause, repetitions, stuttering, and prolongation~\cite{filledpause,repairAndRepetition,prolongation}.

\textbf{(2) Injections:}
An interjection is a word or expression that expresses a spontaneous feeling or reaction~\cite{interjectionsLin}.
We categorize them into doubt, response, surprise, positive feedback, reminder, realization, sigh, coquetry and snort based on pragmatic functions.

\textbf{(3) Non-speech sound:}
Non-speech sounds include all recognizable verbal but non-speech sounds, for instance laughter~\cite{spontaneousClassify}.
We further subdivide laughter into the smile, cachinnation, wry smile, awkward laughter, scoff and involuntary laughter.


Our model focuses on the modeling of the 19 spontaneous behaviors mentioned above, which are explicitly annotated in the transcripts of the spontaneous style corpus\footnote{More details on the definitions, lexical features and acoustic characteristics of spontaneous behaviors can be found on: 
\href{https://thuhcsi.github.io/interspeech2024-SponLMTTS}{https://thuhcsi.github.io/interspeech2024-SponLMTTS}\label{demo}}.

\begin{figure}[!tb]
	\centering
	\includegraphics[width=0.6 \linewidth, height=0.7\linewidth]{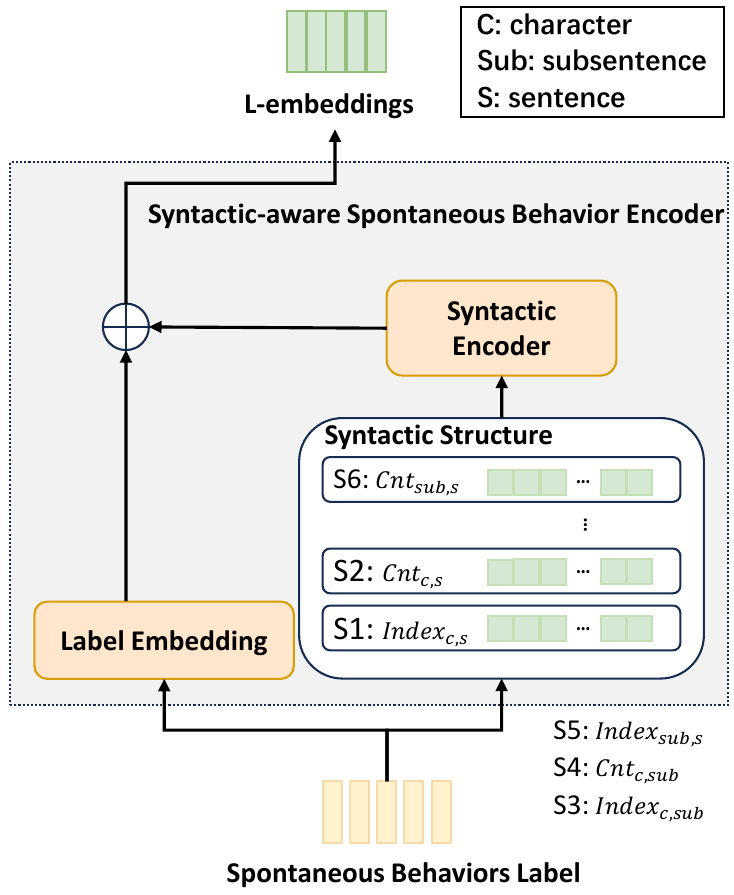}
	\caption{The architecture of syntactic-aware spontaneous behavior encoder. The ${Index_{a,b}}$ represents ${a}$'s index in ${b}$, the ${Cnt_{a,b}}$ represents number of ${a}$ in the ${b}$. Subsentences are separated by punctuation.}
	\label{fig:Lencoder}
\end{figure}

\vspace{-0.2cm}
\subsection{Syntactic-aware Spontaneous Behavior Modeling}
\vspace{-0.1cm}

To model the above diverse spontaneous behaviors, we propose a syntactic-aware spontaneous behavior encoder consisting of a label embedding and a syntactic encoder, as shown in Figure. \ref{fig:Lencoder}.

The character-level spontaneous label sequence is expanded to the phoneme-level by simple repetition, and the representation of spontaneous behavior is obtained through a label embedding.
The syntactic positions of spontaneous behaviors correspond to their various pragmatic functions, 
which leads us to extract the syntactic information of spontaneous behavior. 
We obtain the syntactic structure for each spontaneous behavior label at the character level, while parts without spontaneous behavior are represented by zero.

After extending the syntactic information to the phoneme level and processing it with a syntactic encoder consisting of two linear layers and ReLU, we combine the resulting syntactic representation with the label embedding output to obtain the syntactic-aware spontaneous behavior embeddings (L-embeddings).
The L-embeddings are added to the text embeddings, ensuring that each text token is accurately added with the corresponding spontaneous behavior, and making better projections of spontaneous representation into the text feature space during the fine-tuning process.

In addition, a NAR label predictor consisting of transformer decoders is used to predict the spontaneous behavior sequence from text embeddings, which can enhance the stability of the prediction results. The predicted spontaneous labels enable the model to synthesize semantically coherent spontaneous behavior without explicit labels.

\vspace{-0.2cm}
\subsection{Spontaneous Prosody Modeling}
\vspace{-0.1cm}
To simulate the varied prosody features in spontaneous speech, we introduce a spontaneous prosody extractor to extract fine-grained prosody representations associated with spontaneous behavior from mel-spectrograms, and it guides the prediction explicitly.
In addition, a LM-based prosody predictor is employed to generate spontaneous prosody representations (P-embeddings), which are added to the text embeddings as conditions for the acoustic decoder.

\vspace{-0.2cm}
\subsubsection{Spontaneous Prosody Extractor}
\begin{figure}[!tb]
	\centering
	\includegraphics[width=0.8 \linewidth, height=0.7\linewidth]{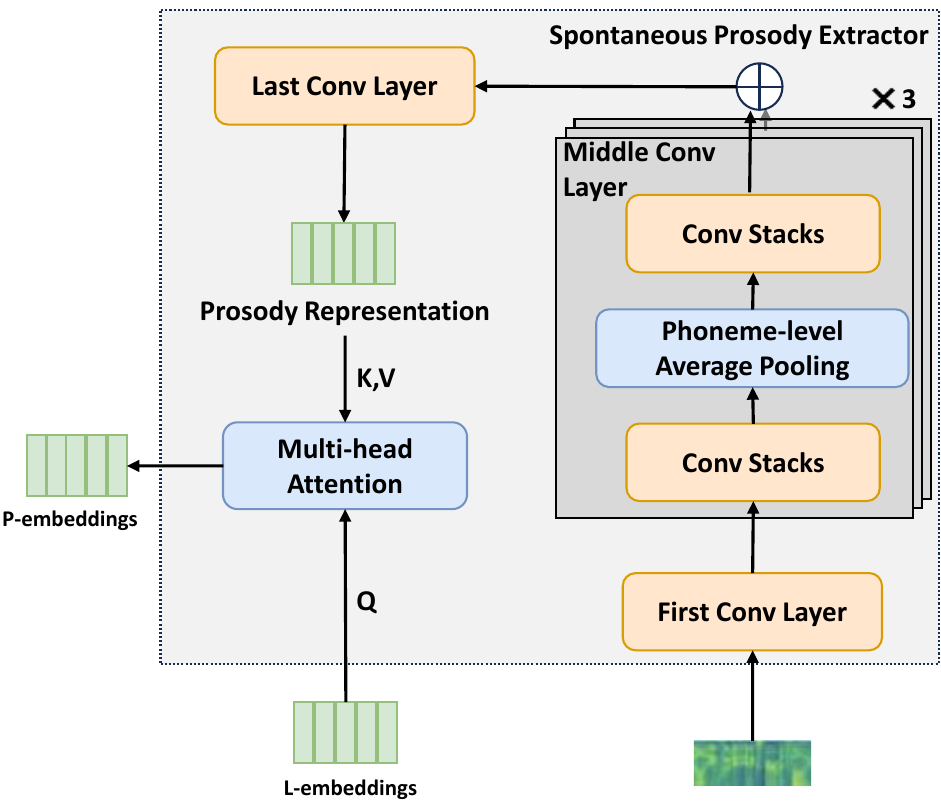}
	\caption{The architecture of Spontaneous Prosody Extractor}
	\label{fig:PE}
        \vspace{-0.3cm}
\end{figure}

\vspace{-0.1cm}
The spontaneous prosody extractor has three convolution layers and a multi-head attention mechanism, as described in Figure. \ref{fig:PE}.
The first convolution layer condenses mel-spectrograms into frame-level hidden states. Then, three middle layers, including two convolution stacks and an average pooling layer, generate intermediate phoneme-level representations based on phoneme boundaries. The final convolution layer combines and processes them to produce fine-grained prosody representation.

Spontaneous behavior significantly influences the prosody of spontaneous speech, so we obtain spontaneous prosody embeddings by merging prosody and spontaneous behaviors information. We use a multi-head attention mechanism, with L-embeddings as the query and prosody representation as the key and value, to understand the prosody selection of each spontaneous label, generating P-embeddings.

\vspace{-0.2cm}
\subsubsection{LM-based Prosody Predictor}
\vspace{-0.1cm}
Since spontaneous prosody is related to text, spontaneous behaviors, and syntactic structure information,
during the inference process, we propose an autoregressive prosody predictor based on a language model consisting of transformer
decoders that takes in these information and predicts P-embeddings.
The AR predictor can capture both local and global dependencies for prosody modeling, helping the acoustic decoder generate expressive spontaneous speech.

\vspace{-0.2cm}
\subsection{Pre-training and Fine-tuning}
\vspace{-0.1cm}
Firstly, we pre-train the backbone model described in section 2.1 on a large-scale dataset, enabling it to acquire strong semantic understanding and the ability to generate expressive speech.
During the fine-tuning stage, we use the pre-trained acoustic model to guide the training of the other modules, allowing the extracted spontaneous behavior and prosody representations to be better projected into the feature space.

Specifically, we divide the fine-tuning stage into three steps:
1) We jointly train the backbone model, the spontaneous behavior encoder, the label predictor, and the spontaneous prosody extractor, excluding the audio codec.
The loss function is shown in equation \ref{eq:loss_function}, where ${C}$ represents the acoustic tokens, ${L}$ represents the spontaneous labels, and ${\lambda}$ is ${0.1}$.
2) We use the trained spontaneous prosody extractor as a teaching model and train the prosody predictor separately. The MSE loss between the predicted and extracted P-embeddings is the loss function.
3) Considering the losses in prosody prediction, we jointly train the prosody predictor and the acoustic model with a lower learning rate while freezing the spontaneous prosody extractor, in order to further improve the naturalness of the synthesized speech.

\vspace{-0.5cm}
\begin{equation}
\mathcal{L} = \mathcal{L}_{\text{ce}}(C_{gt},C_{predict}) + \lambda \mathcal{L}_{\text{ce}}(L_{gt},L_{predict})
\label{eq:loss_function}
\end{equation}

\section{Experiments}



\vspace{-0.1cm}
\subsection{Basic Setup}
\vspace{-0.1cm}
We conduct our experiments on two Mandarin corpora.
For pre-training, we use the open-source corpus WenetSpeech~\cite{Wenet} that contains 10k hours of multi-domain speeches.
For fine-tuning, we use a high-quality internal dataset, which includes 5.4 hours of spontaneous speech consisting of 4968 records from a native Mandarin female speaker naturally pronouncing within the context of specific conversation texts.
It contains 19 behavior annotations as described in section 2.2, for a total of 5907 spontaneous labels.
We randomly selected 200 spontaneous records as test set.
A pre-trained neural audio codec model, EnCodec\footnote{Implemented based on: \href{https://github.com/facebookresearch/encodec}{https://github.com/facebookresearch/encodec}\label{encodec}}\cite{encodec}, was used to encode the original waveform at 24kHz sampling rate and reconstruct the waveform based on the predicted acoustic tokens.

In our implementation, the acoustic decoder follows the setting of VALL-E. The label predictor and prosody predictor consist of 3 layers of transformer blocks.
The spontaneous prosody extractor consists of 8 convolution layers with a kernel size of 5.
We pre-train the backbone model for 40 epochs on 8 NVIDIA A100 GPUs.
During the fine-tuning stage,
we train 30k iterations for the first train step, then 20k for the second train step and 16k for the final train step.
The ScaledAdam~\cite{ScaledAdam} optimizer is adopted with $\beta_1=0.9$, $\beta_2=0.95$.
\vspace{-0.1cm}
\subsection{Compared Methods}
\vspace{-0.1cm}
We compare four models for spontaneous speech synthesis.

\textbf{FastSpeech 2}
A vanilla FastSpeech 2 which is trained on the spontaneous corpus, dose no explicitly behavior modeling.

\textbf{VALL-E}
An open-source implementation
of VALL-E~\cite{valle}. We trained the model in two Mandarin corpora and used it as our baseline model, as described in section 2.1.

\textbf{Base-L}
The VALL-E with the syntactic-aware spontaneous behavior modeling, excludes the prosody representations.

\textbf{Proposed}
The proposed model
that considers both syntactic-aware spontaneous behavior modeling and spontaneous prosody modeling based on VALL-E.
\vspace{-0.2cm}
\subsection{Subjective Evaluation}
\vspace{-0.1cm}
To evaluate the ability of the model to generate natural spontaneous speech, we conducted two mean opinion score (MOS) tests.
The first test assesses the prosody naturalness of the overall speech (PN-MOS) and the second test focuses only on the naturalness of the spontaneous behavior (LN-MOS), and we both provide explicit labels for control.
We randomly selected 20 sentences from the test set and invited 25 native Mandarin speakers to rate the speeches from 1 to 5.
As shown in Table \ref{tab:mos},
the proposed method significantly outperforms other models in both prosody naturalness and spontaneous behavior naturalness.
The performance of VALL-E significantly surpasses that of FastSpeech 2, demonstrating the capability of large-scale language models in synthesizing natural speech.
Compared to the baseline, Base-L with the introduction of spontaneous behavior modeling significantly improves speech naturalness, particularly for the naturalness of spontaneous behavior.
Our proposed method achieves 4.09 in PN-MOS and 4.046 in LN-MOS, indicating that our model is capable of synthesizing spontaneous speech that is closer to human performance.

\begin{figure}[!htb]
	\centering
	\includegraphics[width=0.8\linewidth, height=0.2\linewidth]{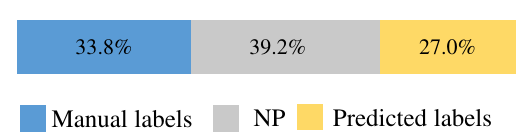}
	\caption{Subjective preference test results on the preference for spontaneous style. Both are generated from the proposed method. NP represents no preference.}
	\label{fig:abx1}
\end{figure}

In addition, a subjective preference test is administered to ask subjects for their preferences regarding the spontaneous style of a pair of sentences.
We conducted a comparison between the speech generated using manual labels and the speech generated from predicted labels.
As shown in Figure~\ref{fig:abx1},
the result for manual labels is only 6.8\% higher than that for predicted labels, and there is 39.2\% no preference result, suggesting that the use of a label predictor allows the model to predict reasonable spontaneous behaviors from the text information.
\vspace{-0.2cm}
\subsection{Objective Evaluation}
\vspace{-0.1cm}
In order to objectively measure the naturalness of the synthesised speech, we computed Mel Cepstral Distortion (MCD)~\cite{mcd}. To account for potential differences in the lengths of predicted and ground-truth speech, we utilized dynamic time warping (DTW) to establish the alignment between the two mel-spectrograms. Next, we determined the minimum
MCD
by aligning the two mel-spectrograms.
The evaluation results of different models on the test set is shown in Table~\ref{tab:mos}.
The results show that our proposed model outperforms other models, indicating its ability to synthesize spontaneous speech.

\vspace{0.15cm}
\begin{table}[th]\footnotesize
\renewcommand{\arraystretch}{1.0}
  \caption{The MOS on naturalness of different models with 95\% confidence intervals and the MCD results.}
  \vspace{-0.2cm}
  \label{tab:mos}
  \centering
  \begin{tabular}{l|c|c|c} 
    \toprule
    \textbf{Models} &\textbf{PN-MOS $\uparrow$}  &\textbf{LN-MOS $\uparrow$} &\textbf{MCD $\downarrow$} \\
    \midrule
    FastSpeech 2 &$2.606\pm0.106$ & $2.536\pm0.097$ ~~~  & $5.355$ ~~~\\
    VALL-E &$3.302\pm0.099$ & $3.324\pm0.098$ ~~~  & $5.291$ ~~~\\
    Base-L &$3.792\pm0.084$ & $3.898\pm0.084$ ~~~  & $4.961$ ~~~\\
    Proposed &$\mathbf{4.090\pm0.073}$ & $\mathbf{4.046\pm0.737}$ ~~~  & $\mathbf{4.879}$ ~~~\\
    \bottomrule
  \end{tabular}
\end{table}

\vspace{-0.6cm}
\subsection{Ablation Study}
\vspace{-0.1cm}
Two ablation studies are conducted to demonstrate the efficacy of several techniques utilized in our proposed model, including the use of spontaneous prosody modeling and spontaneous behavior modeling.
The Comparative Mean Opinion Score (CMOS) was used to compare synthetic speech in terms of the prosody naturalness of the overall speech (PN) and the naturalness of spontaneous behavior (LN).
The results are shown in Table \ref{tab:cmos}.
Eliminating the use of spontaneous prosody results in $-0.320$ CMOS on the prosody naturalness of the speech.
This demonstrates that the addition of spontaneous prosody representations greatly improved the naturalness of the synthesised speech. 
Moreover, we find that not explicitly model spontaneous behavior resulted in $-0.408$ CMOS. This indicates that explicit label control enhances the model's ability to simulate spontaneous behaviour.
\vspace{-0.4cm}
\vspace{0.5cm}
\begin{table}[th]\footnotesize
\renewcommand{\arraystretch}{1.0}
  \caption{CMOS comparison for ablation study.}
  \label{tab:cmos}
  \centering
  \begin{tabular}{l|c} 
    \toprule
    \textbf{Modes} &\textbf{CMOS} \\
    \midrule
    Proposed & $0$ ~~~ \\
    \quad -spontaneous prosody modeling (PN) & $-0.320$ ~~~ \\
    \quad -spontaneous behavior modeling (LN) & $-0.408$ ~~~ \\
    \bottomrule
  \end{tabular}
\end{table}

\vspace{-0.3cm}
\begin{figure}[!htb]
	\centering
	\includegraphics[width=1\linewidth, height=0.4\linewidth]{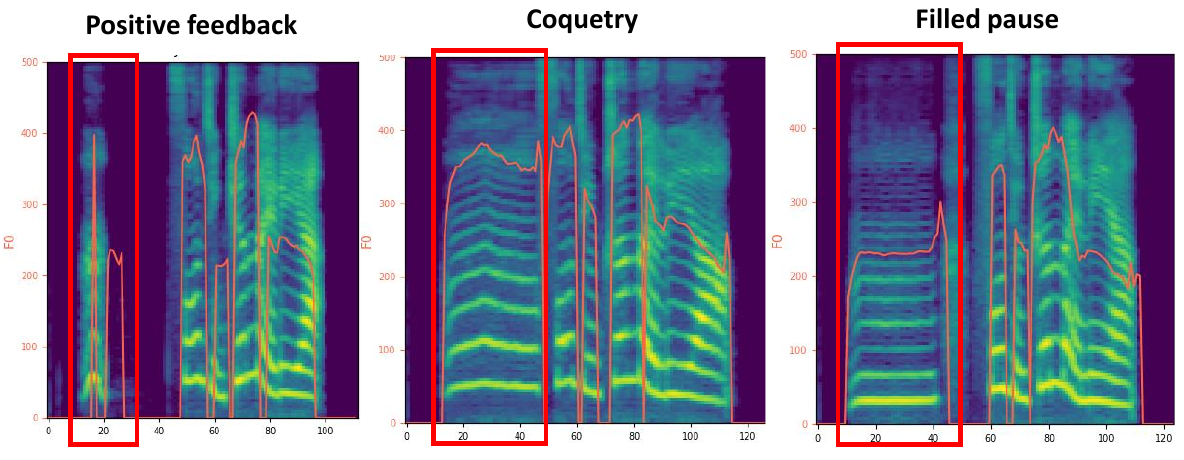}
	\caption{The mel-spectrograms and pitch contours of speech synthesized by the proposed model. The text means ``um, the scenery is really beautiful.'' and different labels are added to ``um'' highlighed by the red box.}
	\label{fig:caseStudy}
\end{figure}

\subsection{Case Study}
\vspace{-0.1cm}
In order to explore the ability of the model to control spontaneous phenomena, a case study is conducted to synthesize the same utterance with different labels.
As shown in Figure~\ref{fig:caseStudy}, the mel-spectrograms, pitch contours, and durations of these speeches are significantly different.
The duration of positive feedback is the shortest, the pitch of coquetry is higher, while the tone of filled pause is lower, and the following speech is also relatively low-pitched.
The results of the case study demonstrate that our model can effectively control spontaneous behavior in spontaneous speech, and affect the stress, pitch, and duration of the synthesized speech.
Some audio samples are provided for listening\footnote{Sample: 
\href{https://thuhcsi.github.io/interspeech2024-SponLMTTS}{https://thuhcsi.github.io/interspeech2024-SponLMTTS}\label{demo}}.

\section{Conclusions}
In this paper, we propose a novel spontaneous speech synthesis system based on language models that can learn powerful
semantic understanding and diverse speech expressions from a massive amount of data.
We incorporate explicit spontaneous behavior modeling and use fine-grained spontaneous prosody representations to enhance the model's ability to synthesize spontaneous speech.
Experimental results show 
our proposed method significantly outperforms the baseline method in terms of prosody naturalness and spontaneous behavior naturalness.

\newpage
\section{Acknowledgement}
This work is supported by National Natural Science Foundation of China (62076144), National Social Science Foundation of China (13\&ZD189), Shenzhen Science and Technology Program (WDZC20220816140515001) and Tencent Rhino-Bird Focused Research Program (RBFR2023015).
\bibliographystyle{IEEEtran}
\bibliography{references}

\end{document}